\documentstyle[aps,multicol,latexsym]{revtex}

\newcommand{\pr}{\partial}
\newcommand{\ub}{\overline{u}}
\newcommand{\pb}{\overline{p}}
\newcommand{\uub}{{\overline{{\bf{u}}}}}
\newcommand{\uu}{{\bf{u}}}
\newcommand{\be}{\begin{equation}}
\newcommand{\en}{\end{equation}}

\newcommand{\bea}{\begin{eqnarray}}
\newcommand{\ena}{\end{eqnarray}}

\begin{document}

\def\PsfigVersion{1.10}
\def\setDriver{\DvipsDriver} 
\ifx\undefined\psfig\else \fi
%

\let\LaTeXAtSign=\@
\let\@=\relax
\edef\psfigRestoreAt{\catcode`\@=\number\catcode`@\relax}
\catcode`\@=11\relax
\newwrite\@unused
\def\ps@typeout#1{{\let\protect\string\immediate\write\@unused{#1}}}

\def\DvipsDriver{
	\ps@typeout{psfig/tex \PsfigVersion -dvips}
\def\PsfigSpecials{\DvipsSpecials} 	\def\ps@dir{/}
\def\ps@predir{} }
\def\OzTeXDriver{
	\ps@typeout{psfig/tex \PsfigVersion -oztex}
	\def\PsfigSpecials{\OzTeXSpecials}
	\def\ps@dir{:}
	\def\ps@predir{:}
	\catcode`\^^J=5
}


\def\figurepath{./:}
\def\psfigurepath#1{\edef\figurepath{#1:}}

\def\DoPaths#1{\expandafter\EachPath#1\stoplist}
\def\leer{}
\def\EachPath#1:#2\stoplist{
  \ExistsFile{#1}{\SearchedFile}
  \ifx#2\leer
  \else
    \expandafter\EachPath#2\stoplist
  \fi}
%
%
\def\ps@dir{/}
\def\ExistsFile#1#2{%
   \openin1=\ps@predir#1\ps@dir#2
   \ifeof1
       \closein1
   \else
       \closein1
        \ifx\ps@founddir\leer
           \edef\ps@founddir{#1}
        \fi
   \fi}
%
%
\def\get@dir#1{%
  \def\ps@founddir{}
  \def\SearchedFile{#1}
  \DoPaths\figurepath
}

%
%
\def\@nnil{\@nil}
\def\@empty{}
\def\@psdonoop#1\@@#2#3{}
\def\@psdo#1:=#2\do#3{\edef\@psdotmp{#2}\ifx\@psdotmp\@empty \else
    \expandafter\@psdoloop#2,\@nil,\@nil\@@#1{#3}\fi}
\def\@psdoloop#1,#2,#3\@@#4#5{\def#4{#1}\ifx #4\@nnil \else
       #5\def#4{#2}\ifx #4\@nnil \else#5\@ipsdoloop #3\@@#4{#5}\fi\fi}
\def\@ipsdoloop#1,#2\@@#3#4{\def#3{#1}\ifx #3\@nnil 
       \let\@nextwhile=\@psdonoop \else
      #4\relax\let\@nextwhile=\@ipsdoloop\fi\@nextwhile#2\@@#3{#4}}
\def\@tpsdo#1:=#2\do#3{\xdef\@psdotmp{#2}\ifx\@psdotmp\@empty \else
    \@tpsdoloop#2\@nil\@nil\@@#1{#3}\fi}
\def\@tpsdoloop#1#2\@@#3#4{\def#3{#1}\ifx #3\@nnil 
       \let\@nextwhile=\@psdonoop \else
      #4\relax\let\@nextwhile=\@tpsdoloop\fi\@nextwhile#2\@@#3{#4}}
%
\ifx\undefined\fbox
\newdimen\fboxrule
\newdimen\fboxsep
\newdimen\ps@tempdima
\newbox\ps@tempboxa
\fboxsep = 3pt
\fboxrule = .4pt
\long\def\fbox#1{\leavevmode\setbox\ps@tempboxa\hbox{#1}\ps@tempdima\fboxrule
    \advance\ps@tempdima \fboxsep \advance\ps@tempdima \dp\ps@tempboxa
   \hbox{\lower \ps@tempdima\hbox
  {\vbox{\hrule height \fboxrule
          \hbox{\vrule width \fboxrule \hskip\fboxsep
          \vbox{\vskip\fboxsep \box\ps@tempboxa\vskip\fboxsep}\hskip 
                 \fboxsep\vrule width \fboxrule}
                 \hrule height \fboxrule}}}}
\fi
%
%
\newread\ps@stream
\newif\ifnot@eof       
\newif\if@noisy        
\newif\if@atend        
\newif\if@psfile       
%
%
{\catcode`\%=12\global\gdef\epsf@start{
\def\epsf@PS{PS}
\def\epsf@getbb#1{%
%
%
\openin\ps@stream=\ps@predir#1
\ifeof\ps@stream\ps@typeout{Error, File #1 not found}\else
%
%
   {\not@eoftrue \chardef\other=12
    \def\do##1{\catcode`##1=\other}\dospecials \catcode`\ =10
    \loop
       \if@psfile
	  \read\ps@stream to \epsf@fileline
       \else{
	  \obeyspaces
          \read\ps@stream to \epsf@tmp\global\let\epsf@fileline\epsf@tmp}
       \fi
       \ifeof\ps@stream\not@eoffalse\else
%
%
       \if@psfile\else
       \expandafter\epsf@test\epsf@fileline:. \\%
       \fi
%
%
          \expandafter\epsf@aux\epsf@fileline:. \\%
       \fi
   \ifnot@eof\repeat
   }\closein\ps@stream\fi}%
%
%
\long\def\epsf@test#1#2#3:#4\\{\def\epsf@testit{#1#2}
			\ifx\epsf@testit\epsf@start\else
\ps@typeout{Warning! File does not start with `\epsf@start'.  It may not be a PostScript file.}
			\fi
			\@psfiletrue} 
%
%
{\catcode`\%=12\global\let\epsf@percent=
%
%
%
\long\def\epsf@aux#1#2:#3\\{\ifx#1\epsf@percent
   \def\epsf@testit{#2}\ifx\epsf@testit\epsf@bblit
	\@atendfalse
        \epsf@atend #3 . \\%
	\if@atend	
	   \if@verbose{
		\ps@typeout{psfig: found `(atend)'; continuing search}
	   }\fi
        \else
        \epsf@grab #3 . . . \\%
        \not@eoffalse
        \global\no@bbfalse
        \fi
   \fi\fi}%
%
%
\def\epsf@grab #1 #2 #3 #4 #5\\{%
   \global\def\epsf@llx{#1}\ifx\epsf@llx\empty
      \epsf@grab #2 #3 #4 #5 .\\\else
   \global\def\epsf@lly{#2}%
   \global\def\epsf@urx{#3}\global\def\epsf@ury{#4}\fi}%
%
%
\def\epsf@atendlit{(atend)} 
\def\epsf@atend #1 #2 #3\\{%
   \def\epsf@tmp{#1}\ifx\epsf@tmp\empty
      \epsf@atend #2 #3 .\\\else
   \ifx\epsf@tmp\epsf@atendlit\@atendtrue\fi\fi}


\chardef\psletter = 11 
\chardef\other = 12

\newif \ifdebug 
\newif\ifc@mpute 
\c@mputetrue 

\let\then = \relax
\def\r@dian{pt }
\let\r@dians = \r@dian
\let\dimensionless@nit = \r@dian
\let\dimensionless@nits = \dimensionless@nit
\def\internal@nit{sp }
\let\internal@nits = \internal@nit
\newif\ifstillc@nverging
\def \Mess@ge #1{\ifdebug \then \message {#1} \fi}

{ 
	\catcode `\@ = \psletter
	\gdef \nodimen {\expandafter \n@dimen \the \dimen}
	\gdef \term #1 #2 #3%
	       {\edef \t@ {\the #1}
		\edef \t@@ {\expandafter \n@dimen \the #2\r@dian}%
		\t@rm {\t@} {\t@@} {#3}%
	       }
	\gdef \t@rm #1 #2 #3%
	       {{%
		\count 0 = 0
		\dimen 0 = 1 \dimensionless@nit
		\dimen 2 = #2\relax
		\Mess@ge {Calculating term #1 of \nodimen 2}%
		\loop
		\ifnum	\count 0 < #1
		\then	\advance \count 0 by 1
			\Mess@ge {Iteration \the \count 0 \space}%
			\Multiply \dimen 0 by {\dimen 2}%
			\Mess@ge {After multiplication, term = \nodimen 0}%
			\Divide \dimen 0 by {\count 0}%
			\Mess@ge {After division, term = \nodimen 0}%
		\repeat
		\Mess@ge {Final value for term #1 of 
				\nodimen 2 \space is \nodimen 0}%
		\xdef \Term {#3 = \nodimen 0 \r@dians}%
		\aftergroup \Term
	       }}
	\catcode `\p = \other
	\catcode `\t = \other
	\gdef \n@dimen #1pt{#1} 
}

\def \Divide #1by #2{\divide #1 by #2} 

\def \Multiply #1by #2
       {{
	\count 0 = #1\relax
	\count 2 = #2\relax
	\count 4 = 65536
	\Mess@ge {Before scaling, count 0 = \the \count 0 \space and
			count 2 = \the \count 2}%
	\ifnum	\count 0 > 32767 
	\then	\divide \count 0 by 4
		\divide \count 4 by 4
	\else	\ifnum	\count 0 < -32767
		\then	\divide \count 0 by 4
			\divide \count 4 by 4
		\else
		\fi
	\fi
	\ifnum	\count 2 > 32767 
	\then	\divide \count 2 by 4
		\divide \count 4 by 4
	\else	\ifnum	\count 2 < -32767
		\then	\divide \count 2 by 4
			\divide \count 4 by 4
		\else
		\fi
	\fi
	\multiply \count 0 by \count 2
	\divide \count 0 by \count 4
	\xdef \product {#1 = \the \count 0 \internal@nits}%
	\aftergroup \product
       }}

\def\r@duce{\ifdim\dimen0 > 90\r@dian \then   
		\multiply\dimen0 by -1
		\advance\dimen0 by 180\r@dian
		\r@duce
	    \else \ifdim\dimen0 < -90\r@dian \then  
		\advance\dimen0 by 360\r@dian
		\r@duce
		\fi
	    \fi}

\def\Sine#1%
       {{%
	\dimen 0 = #1 \r@dian
	\r@duce
	\ifdim\dimen0 = -90\r@dian \then
	   \dimen4 = -1\r@dian
	   \c@mputefalse
	\fi
	\ifdim\dimen0 = 90\r@dian \then
	   \dimen4 = 1\r@dian
	   \c@mputefalse
	\fi
	\ifdim\dimen0 = 0\r@dian \then
	   \dimen4 = 0\r@dian
	   \c@mputefalse
	\fi
	\ifc@mpute \then
		\divide\dimen0 by 180
		\dimen0=3.141592654\dimen0
		\dimen 2 = 3.1415926535897963\r@dian 
		\divide\dimen 2 by 2 
		\Mess@ge {Sin: calculating Sin of \nodimen 0}%
		\count 0 = 1 
		\dimen 2 = 1 \r@dian 
		\dimen 4 = 0 \r@dian 
		\loop
			\ifnum	\dimen 2 = 0 
			\then	\stillc@nvergingfalse 
			\else	\stillc@nvergingtrue
			\fi
			\ifstillc@nverging 
			\then	\term {\count 0} {\dimen 0} {\dimen 2}%
				\advance \count 0 by 2
				\count 2 = \count 0
				\divide \count 2 by 2
				\ifodd	\count 2 
				\then	\advance \dimen 4 by \dimen 2
				\else	\advance \dimen 4 by -\dimen 2
				\fi
		\repeat
	\fi		
			\xdef \sine {\nodimen 4}%
       }}

\def\Cosine#1{\ifx\sine\UnDefined\edef\Savesine{\relax}\else
		             \edef\Savesine{\sine}\fi
	{\dimen0=#1\r@dian\advance\dimen0 by 90\r@dian
	 \Sine{\nodimen 0}
	 \xdef\cosine{\sine}
	 \xdef\sine{\Savesine}}}	      

\def\psdraft{
	\def\@psdraft{0}
}
\def\psfull{
	\def\@psdraft{100}
}

\psfull

\newif\if@scalefirst
\def\psscalefirst{\@scalefirsttrue}
\def\psrotatefirst{\@scalefirstfalse}
\psrotatefirst

\newif\if@draftbox
\def\psnodraftbox{
	\@draftboxfalse
}
\def\psdraftbox{
	\@draftboxtrue
}
\@draftboxtrue

\newif\if@prologfile
\newif\if@postlogfile
\def\pssilent{
	\@noisyfalse
}
\def\psnoisy{
	\@noisytrue
}
\psnoisy
\newif\if@bbllx
\newif\if@bblly
\newif\if@bburx
\newif\if@bbury
\newif\if@height
\newif\if@width
\newif\if@rheight
\newif\if@rwidth
\newif\if@angle
\newif\if@clip
\newif\if@verbose
\def\@p@@sclip#1{\@cliptrue}
\newif\if@decmpr
\def\@p@@sfigure#1{\def\@p@sfile{null}\def\@p@sbbfile{null}\@decmprfalse
   \openin1=\ps@predir#1
   \ifeof1
	\closein1
	\get@dir{#1}
	\ifx\ps@founddir\leer
		\openin1=\ps@predir#1.bb
		\ifeof1
			\closein1
			\get@dir{#1.bb}
			\ifx\ps@founddir\leer
				\ps@typeout{Can't find #1 in \figurepath}
			\else
				\@decmprtrue
				\def\@p@sfile{\ps@founddir\ps@dir#1}
				\def\@p@sbbfile{\ps@founddir\ps@dir#1.bb}
			\fi
		\else
			\closein1
			\@decmprtrue
			\def\@p@sfile{#1}
			\def\@p@sbbfile{#1.bb}
		\fi
	\else
		\def\@p@sfile{\ps@founddir\ps@dir#1}
		\def\@p@sbbfile{\ps@founddir\ps@dir#1}
	\fi
   \else
	\closein1
	\def\@p@sfile{#1}
	\def\@p@sbbfile{#1}
   \fi
}
\def\@p@@sfile#1{\@p@@sfigure{#1}}
\def\@p@@sbbllx#1{
		\@bbllxtrue
		\dimen100=#1
		\edef\@p@sbbllx{\number\dimen100}
}
\def\@p@@sbblly#1{
		\@bbllytrue
		\dimen100=#1
		\edef\@p@sbblly{\number\dimen100}
}
\def\@p@@sbburx#1{
		\@bburxtrue
		\dimen100=#1
		\edef\@p@sbburx{\number\dimen100}
}
\def\@p@@sbbury#1{
		\@bburytrue
		\dimen100=#1
		\edef\@p@sbbury{\number\dimen100}
}
\def\@p@@sheight#1{
		\@heighttrue
		\dimen100=#1
   		\edef\@p@sheight{\number\dimen100}
}
\def\@p@@swidth#1{
		\@widthtrue
		\dimen100=#1
		\edef\@p@swidth{\number\dimen100}
}
\def\@p@@srheight#1{
		\@rheighttrue
		\dimen100=#1
		\edef\@p@srheight{\number\dimen100}
}
\def\@p@@srwidth#1{
		\@rwidthtrue
		\dimen100=#1
		\edef\@p@srwidth{\number\dimen100}
}
\def\@p@@sangle#1{
		\@angletrue
		\edef\@p@sangle{#1} 
}
\def\@p@@ssilent#1{ 
		\@verbosefalse
}
\def\@p@@sprolog#1{\@prologfiletrue\def\@prologfileval{#1}}
\def\@p@@spostlog#1{\@postlogfiletrue\def\@postlogfileval{#1}}
\def\@cs@name#1{\csname #1\endcsname}
\def\@setparms#1=#2,{\@cs@name{@p@@s#1}{#2}}
%
%
\def\ps@init@parms{
		\@bbllxfalse \@bbllyfalse
		\@bburxfalse \@bburyfalse
		\@heightfalse \@widthfalse
		\@rheightfalse \@rwidthfalse
		\def\@p@sbbllx{}\def\@p@sbblly{}
		\def\@p@sbburx{}\def\@p@sbbury{}
		\def\@p@sheight{}\def\@p@swidth{}
		\def\@p@srheight{}\def\@p@srwidth{}
		\def\@p@sangle{0}
		\def\@p@sfile{} \def\@p@sbbfile{}
		\def\@p@scost{10}
		\def\@sc{}
		\@prologfilefalse
		\@postlogfilefalse
		\@clipfalse
		\if@noisy
			\@verbosetrue
		\else
			\@verbosefalse
		\fi
}
%
%
\def\parse@ps@parms#1{
	 	\@psdo\@psfiga:=#1\do
		   {\expandafter\@setparms\@psfiga,}}
%
%
\newif\ifno@bb
\def\bb@missing{
	\if@verbose{
		\ps@typeout{psfig: searching \@p@sbbfile \space  for bounding box}
	}\fi
	\no@bbtrue
	\epsf@getbb{\@p@sbbfile}
        \ifno@bb \else \bb@cull\epsf@llx\epsf@lly\epsf@urx\epsf@ury\fi
}	
\def\bb@cull#1#2#3#4{
	\dimen100=#1 bp\edef\@p@sbbllx{\number\dimen100}
	\dimen100=#2 bp\edef\@p@sbblly{\number\dimen100}
	\dimen100=#3 bp\edef\@p@sbburx{\number\dimen100}
	\dimen100=#4 bp\edef\@p@sbbury{\number\dimen100}
	\no@bbfalse
}
\newdimen\p@intvaluex
\newdimen\p@intvaluey
\def\rotate@#1#2{{\dimen0=#1 sp\dimen1=#2 sp
		  \global\p@intvaluex=\cosine\dimen0
		  \dimen3=\sine\dimen1
		  \global\advance\p@intvaluex by -\dimen3
		  \global\p@intvaluey=\sine\dimen0
		  \dimen3=\cosine\dimen1
		  \global\advance\p@intvaluey by \dimen3
		  }}
\def\compute@bb{
		\no@bbfalse
		\if@bbllx \else \no@bbtrue \fi
		\if@bblly \else \no@bbtrue \fi
		\if@bburx \else \no@bbtrue \fi
		\if@bbury \else \no@bbtrue \fi
		\ifno@bb \bb@missing \fi
		\ifno@bb \ps@typeout{FATAL ERROR: no bb supplied or found}
			\no-bb-error
		\fi
		%
%
		\count203=\@p@sbburx
		\count204=\@p@sbbury
		\advance\count203 by -\@p@sbbllx
		\advance\count204 by -\@p@sbblly
		\edef\ps@bbw{\number\count203}
		\edef\ps@bbh{\number\count204}
		\if@angle 
			\Sine{\@p@sangle}\Cosine{\@p@sangle}
	        	{\dimen100=\maxdimen\xdef\r@p@sbbllx{\number\dimen100}
					    \xdef\r@p@sbblly{\number\dimen100}
			                    \xdef\r@p@sbburx{-\number\dimen100}
					    \xdef\r@p@sbbury{-\number\dimen100}}
%
                        \def\minmaxtest{
			   \ifnum\number\p@intvaluex<\r@p@sbbllx
			      \xdef\r@p@sbbllx{\number\p@intvaluex}\fi
			   \ifnum\number\p@intvaluex>\r@p@sbburx
			      \xdef\r@p@sbburx{\number\p@intvaluex}\fi
			   \ifnum\number\p@intvaluey<\r@p@sbblly
			      \xdef\r@p@sbblly{\number\p@intvaluey}\fi
			   \ifnum\number\p@intvaluey>\r@p@sbbury
			      \xdef\r@p@sbbury{\number\p@intvaluey}\fi
			   }
			\rotate@{\@p@sbbllx}{\@p@sbblly}
			\minmaxtest
			\rotate@{\@p@sbbllx}{\@p@sbbury}
			\minmaxtest
			\rotate@{\@p@sbburx}{\@p@sbblly}
			\minmaxtest
			\rotate@{\@p@sbburx}{\@p@sbbury}
			\minmaxtest
			\edef\@p@sbbllx{\r@p@sbbllx}\edef\@p@sbblly{\r@p@sbblly}
			\edef\@p@sbburx{\r@p@sbburx}\edef\@p@sbbury{\r@p@sbbury}
		\fi
		\count203=\@p@sbburx
		\count204=\@p@sbbury
		\advance\count203 by -\@p@sbbllx
		\advance\count204 by -\@p@sbblly
		\edef\@bbw{\number\count203}
		\edef\@bbh{\number\count204}
}
%
%
\def\in@hundreds#1#2#3{\count240=#2 \count241=#3
		     \count100=\count240	
		     \divide\count100 by \count241
		     \count101=\count100
		     \multiply\count101 by \count241
		     \advance\count240 by -\count101
		     \multiply\count240 by 10
		     \count101=\count240	
		     \divide\count101 by \count241
		     \count102=\count101
		     \multiply\count102 by \count241
		     \advance\count240 by -\count102
		     \multiply\count240 by 10
		     \count102=\count240	
		     \divide\count102 by \count241
		     \count200=#1\count205=0
		     \count201=\count200
			\multiply\count201 by \count100
		 	\advance\count205 by \count201
		     \count201=\count200
			\divide\count201 by 10
			\multiply\count201 by \count101
			\advance\count205 by \count201
		     \count201=\count200
			\divide\count201 by 100
			\multiply\count201 by \count102
			\advance\count205 by \count201
		     \edef\@result{\number\count205}
}
\def\compute@wfromh{
		\in@hundreds{\@p@sheight}{\@bbw}{\@bbh}
		\edef\@p@swidth{\@result}
}
\def\compute@hfromw{
	        \in@hundreds{\@p@swidth}{\@bbh}{\@bbw}
		\edef\@p@sheight{\@result}
}
\def\compute@handw{
		\if@height 
			\if@width
			\else
				\compute@wfromh
			\fi
		\else 
			\if@width
				\compute@hfromw
			\else
				\edef\@p@sheight{\@bbh}
				\edef\@p@swidth{\@bbw}
			\fi
		\fi
}
\def\compute@resv{
		\if@rheight \else \edef\@p@srheight{\@p@sheight} \fi
		\if@rwidth \else \edef\@p@srwidth{\@p@swidth} \fi
}
%
\def\compute@sizes{
	\compute@bb
	\if@scalefirst\if@angle
	\if@width
	   \in@hundreds{\@p@swidth}{\@bbw}{\ps@bbw}
	   \edef\@p@swidth{\@result}
	\fi
	\if@height
	   \in@hundreds{\@p@sheight}{\@bbh}{\ps@bbh}
	   \edef\@p@sheight{\@result}
	\fi
	\fi\fi
	\compute@handw
	\compute@resv}
\def\OzTeXSpecials{
	\special{empty.ps /@isp {true} def}
	\special{empty.ps \@p@swidth \space \@p@sheight \space
			\@p@sbbllx \space \@p@sbblly \space
			\@p@sbburx \space \@p@sbbury \space
			startTexFig \space }
	\if@clip{
		\if@verbose{
			\ps@typeout{(clip)}
		}\fi
		\special{empty.ps doclip \space }
	}\fi
	\if@angle{
		\if@verbose{
			\ps@typeout{(rotate)}
		}\fi
		\special {empty.ps \@p@sangle \space rotate \space} 
	}\fi
	\if@prologfile
	    \special{\@prologfileval \space } \fi
	\if@decmpr{
		\if@verbose{
			\ps@typeout{psfig: Compression not available
			in OzTeX version \space }
		}\fi
	}\else{
		\if@verbose{
			\ps@typeout{psfig: including \@p@sfile \space }
		}\fi
		\special{epsf=\ps@predir\@p@sfile \space }
	}\fi
	\if@postlogfile
	    \special{\@postlogfileval \space } \fi
	\special{empty.ps /@isp {false} def}
}
\def\DvipsSpecials{
	\special{ps::[begin] 	\@p@swidth \space \@p@sheight \space
			\@p@sbbllx \space \@p@sbblly \space
			\@p@sbburx \space \@p@sbbury \space
			startTexFig \space }
	\if@clip{
		\if@verbose{
			\ps@typeout{(clip)}
		}\fi
		\special{ps:: doclip \space }
	}\fi
	\if@angle
		\if@verbose{
			\ps@typeout{(clip)}
		}\fi
		\special {ps:: \@p@sangle \space rotate \space} 
	\fi
	\if@prologfile
	    \special{ps: plotfile \@prologfileval \space } \fi
	\if@decmpr{
		\openin1=\ps@predir\@p@sfile.gz
		\ifeof1
		        \closein1
			\if@verbose{
				\ps@typeout{psfig: including \@p@sfile.Z \space }
			}\fi
			\special{ps: plotfile "`zcat \@p@sfile.Z" \space }
		\else
                        \closein1
			\if@verbose{
				\ps@typeout{psfig: including \@p@sfile.gz \space }
			}\fi
			\special{ps: plotfile "`gunzip -c \@p@sfile.gz" \space }
		\fi
	}\else{
		\if@verbose{
			\ps@typeout{psfig: including \@p@sfile \space }
		}\fi
		\special{ps: plotfile \@p@sfile \space }
	}\fi
	\if@postlogfile
	    \special{ps: plotfile \@postlogfileval \space } \fi
	\special{ps::[end] endTexFig \space }
}
%
%
\def\psfig#1{\vbox {
	%
	\ps@init@parms
	\parse@ps@parms{#1}
	\compute@sizes
	\ifnum\@p@scost<\@psdraft{
		\PsfigSpecials 
		\vbox to \@p@srheight sp{
			\hbox to \@p@srwidth sp{
				\hss
			}
		\vss
		}
	}\else{
		\if@draftbox{		
			\hbox{\fbox{\vbox to \@p@srheight sp{
			\vss
			\hbox to \@p@srwidth sp{ \hss 
			 \hss }
			\vss
			}}}
		}\else{
			\vbox to \@p@srheight sp{
			\vss
			\hbox to \@p@srwidth sp{\hss}
			\vss
			}
		}\fi

	}\fi
}}
\psfigRestoreAt
\setDriver
\let\@=\LaTeXAtSign

\draft
\title{Regularization modeling for large-eddy simulation
}
\author{Bernard J. Geurts}
\address{Faculty of Mathematical Sciences, J.M. Burgers Center, 
University of Twente\\ P.O. Box 217,
7500 AE Enschede, The Netherlands}
\author{Darryl D. Holm}
\address{Theoretical Division and Center for Nonlinear Studies, Los 
Alamos National Laboratory\\
{\mbox{MS B284 Los Alamos,}} NM 87545, USA}

\maketitle
\begin{abstract}
A new modeling approach for large-eddy simulation (LES) is obtained 
by combining a `regularization principle' with an
explicit filter and its inversion. This regularization approach 
allows a systematic derivation of the implied
subgrid-model, which resolves the closure problem. The central role 
of the filter in LES is fully restored, i.e., both
the interpretation of LES predictions in terms of direct simulation 
results as well as the corresponding subgrid closure
are specified by the filter. The regularization approach is 
illustrated with `Leray-smoothing' of the nonlinear convective
terms. In turbulent mixing the new, implied subgrid model performs 
favorably compared to the dynamic eddy-viscosity procedure. The
model is robust at arbitrarily high Reynolds numbers and correctly 
predicts self-similar turbulent flow development.
\end{abstract}

\newpage

\noindent
{\bf{Corresponding author:}} Bernard J. Geurts

\noindent
{\bf{e-mail:}} b.j.geurts@math.utwente.nl

\noindent
{\bf{tel:}} +31-53-489-4125

\noindent
{\bf{fax:}} +31-53-489-4833


\newpage

Accurate modeling and simulation of turbulent flow is a topic of 
intense ongoing research~\cite{menkatz}. Modern strategies for turbulent
flow are aimed at reducing the dynamical complexity of the underlying 
system of partial differential equations while
reliably predicting the primary flow phenomena. In large-eddy 
simulation (LES) these conflicting requirements are
expressed by coarsening the description on the one hand and subgrid 
modeling on the other hand. The coarsening is
achieved by spatial filtering~\cite{germano} which externally specifies the physical 
detail that will ideally be retained in the LES
solution. Maintaining the dynamical properties of the resolved large 
scales is approached by introducing subgrid modeling
to deal with the closure problem that arises from filtering the 
nonlinear terms.

In the filtering approach to incompressible flow the specification of 
the basic convolution filter $L$ is all that is
required to uniquely define the relation between the unfiltered and 
filtered flow field as well as the closure problem for
the so-called turbulent stress-tensor $\tau_{ij}$. This situation is 
in sharp contrast with actual present-day large-eddy
modeling in which the specification of the subgrid model for 
$\tau_{ij}$ as well as the comparison with reference direct
numerical simulation (DNS) results is performed largely independent 
of the specific choice of the filter $L$.

In this paper we will formulate an alternative approach to large-eddy 
simulations which completely restores the two
central roles of the basic filter $L$, i.e., providing an 
interpretation of LES predictions in terms of filtered DNS
results as well as fully specifying all details of the subgrid model. 
The key elements in this new formulation are a
`regularization principle', a filter $L$ and its (formal) inverse 
operator denoted by $L^{-1}$~\cite{geurts_pof_1997}.

A regularization principle expresses the smoothing of the dynamics 
of the Navier-Stokes equations through a specific proposal for direct
alteration of the nonlinear convective terms. This modeling differs 
significantly from traditional, less direct
approaches, e.g., involving the introduction of additional 
eddy-viscosity contributions~\cite{smagor}. The latter are clearly of a
different physical nature and do not fully do justice to the 
intricate nonlinear transport structure of the filtered
Navier-Stokes equations. The regularization principle gives rise to a 
basic mixed formulation involving both the filtered
and unfiltered solution. Application of $L$ and $L^{-1}$ then allows 
to derive an equivalent representation solely in
terms of the filtered solution. This provides a unique identification 
of the implied subgrid model without any further
external (ad hoc) input or mathematical-physical considerations of 
the closure problem. The regularization modeling
approach is not only theoretically transparent and elegant, but it 
also gives rise to accurate LES predictions. In
particular, we consider the implied subgrid model that arises from 
Leray's regularization principle~\cite{leray}. A comparison between
the Leray model and dynamic subgrid modeling (e.g.,~\cite{vreman_jfm}) will be made for 
turbulent mixing flow, both at moderate and at high
Reynolds numbers.

In the filtering approach one assumes any normalized convolution 
filter $L: u_i \rightarrow \ub_i$ where $\ub_i$ ($u_i$)
denotes the filtered (unfiltered) component of the velocity field in the
$x_i$ direction. Filtering the Navier-Stokes equations yields
\be
\pr_{t}\ub_{i}+\pr_{j}(\ub_{j}\ub_{i})+\pr_{i} 
\pb -\frac{1}{Re}\pr_{jj}\ub_{k}=
-\pr_{j}\tau_{ij}
\label{filteq}
\en
where the turbulent stress tensor 
$\tau_{ij}={\overline{u_{i}u_{j}}}-{\overline{u}}_{i}{\overline{u}}_{j}$ 
represents the
closure problem and $Re$ denotes the Reynolds number. Both the 
relation between $u_{i}$ and $\ub_{i}$ as well as the
properties of $\tau_{ij}$ are fully specified by $L$. In actual 
subgrid modeling for LES, the next step is to introduce a
subgrid model $m_{ij}({\overline{\bf{u}}})$ to approximate 
$\tau_{ij}$. A variety of subgrid models has been proposed to
capture dissipative, dispersive or similarity properties of $\tau_{ij}$.

Many subgrid models are arrived at through a physical or mathematical 
reasoning which is only loosely connected to a
specific filter $L$. As an example, the well-known Smagorinsky model~\cite{smagor} 
is given by
$m_{ij}^{S}=-(C_{S}\Delta)^{2}|S_{ij}({\overline{{\bf{u}}}})|S_{ij}({\overline{{\bf{u}}}})$ 
where the rate of strain
tensor $S_{ij}=\pr_{i}\ub_{j}+\pr_{j}\ub_{i}$ and 
$|S_{ij}|^{2}=S_{ij}S_{ij}/2$. The only explicit reference to the
filter, made in this model, is through the filter-width $\Delta$. In 
actual simulations $\Delta$ is specified in terms of
the grid-spacing $h$ rather than in terms of $L$. Furthermore, the 
Smagorinsky constant $C_{S}$ is determined independent
of $L$, which further reduces any principal role for the filter. The 
situation is comparable for the `tensor-diffusivity'
model $m_{ij}^{TD}=C_{TD} \Delta_k^2 \pr_k \ub_i \pr_k \ub_j$, with 
$\Delta_{k}$ the filter-width in the $x_k$-direction~\cite{clark}.
The coefficient $C_{TD}$ is usually related to the normalized second 
moment $(L(x^{2})-x^{2})/\Delta^{2}$ of the filter
$L$. For various popular filters such as the top-hat or the Gaussian 
filter one finds $C_{TD}=1/12$, i.e., independent of
the actual filter used. The role of the filter is in principle fully 
explicit in Bardina's similarity model
$m_{ij}^{B}={\overline{\ub_{i}\ub_{j}}}-{\overline{\ub}}_{i}{\overline{\ub}}_{j}$ \cite{bardina}.
In actual simulations, however, one
frequently adopts a wider explicit filter or a filter of a different 
type, to enhance smoothing properties of this model \cite{menkatz}.
Moreover, the model is sometimes multiplied by a constant $C_{B}$ 
which is specified independently of any presumed filter \cite{salvbaner}.
Finally, the successful dynamic subgrid modeling requires only the 
explicit specification of the so-called test-filter \cite{germano2}.
To retain the central Germano identity the test-filter
can in principle be chosen independent of $L$, mainly requiring the specification of the filterwidth of the
test-filter relative to $\Delta$. Additional averaging over 
homogeneous directions, `clipping' steps to stabilize
actual simulations, and the fact that the assumed base-models are 
themselves only loosely connected to $L$, also make the
dynamic procedure rather insensitive to the specific assumed filter.

In contrast to these popular LES models, the regularization approach 
involves the introduction of a pair $(L,~L^{-1})$
to fully specify the implied subgrid model as well as the 
interpretation of LES predictions in terms of reference DNS
results. The selection of any other pair $({\cal L},~{\cal L}^{-1})$ 
directly leads to its corresponding DNS
interpretation and the associated subgrid model consistent with the 
regularization principle. This modeling strategy has
a number of important benefits, addressing directly the nonlinear 
convective contributions and requiring no additional `external' 
information such as model coefficients or the width of the test-filter. The regularization principle allows a
transparent modeling in which the modeled system of equations can be 
made to share a number of fundamental properties
with the Navier-Stokes equations, such as transformation symmetries, Kelvin's 
circulation theorem, etc.. The implied subgrid model is quite
simple to implement, with some technical complications arising 
from the construction of an accurate inverse operator
$L^{-1}$.

To illustrate the approach we consider the intuitively appealing and particularly simple Leray regularization in which 
the convective fluxes are replaced by
$\ub_{j}\pr_{j}u_{i}$, i.e., the solution $\uu$ is convected with a 
smoothed velocity $\uub$. Consequently, the nonlinear
effects are reduced by an amount governed by the smoothing properties 
of $L$. The governing equations in the Leray
formulation can be written as \cite{leray}
\begin{equation}
\pr_{j}\ub_{j}=0~~;~~\pr_{t}u_{i}+\ub_{j}\pr_{j}u_{i}+\pr_{i}p-\frac{1}{Re}\pr_{jj}u_{i}=0
\label{leray1}
\end{equation}
Uniqueness and regularity of the solution to these equations have 
been established rigorously \cite{leray}. The Leray
formulation contains the unfiltered Navier-Stokes equations in the 
limiting case $L \rightarrow Id$, e.g., as $\Delta
\rightarrow 0$ ($Id$ denotes the identity). The unfiltered solution 
can readily be eliminated from (\ref{leray1}) by
using the inversion operator $u_{j}=L^{-1}(\ub_{j})$. After some 
calculation (\ref{leray1}) can be written in the same way as the LES `template' (\ref{filteq}) in which $\tau_{ij}$
on the right hand side is replaced by the asymmetric,
filtered similarity-type Leray model $m_{ij}^{L}$ given by:
\begin{equation}
m^{L}_{ij}=L\Big(\ub_{j}L^{-1}(\ub_{i})\Big) 
-\ub_{j}\ub_{i}={\overline{\ub_{j}u_{i}}} -\ub_{j}\ub_{i}
\label{basic_leray}
\end{equation}
This model requires the explicit  application of both $L$ and $L^{-1}$.

In the sequel we consider invertible numerical quadrature 
approximating the top-hat filter. In one dimension the numerical convolution
filtering $\ub=G*u$ corresponds to kernels
\begin{equation}
G(z)=\sum a_{j} \delta(z-z_{j})~~;~~|z_{j}| \leq \Delta/2
\end{equation}
In particular, we consider three-point filters with $a_{0}=1-\alpha$, 
$a_{1}=a_{-1}=\alpha/2$ and $z_{0}=0$,
$z_{1}=-z_{-1}=\Delta/2$. Here we use $\alpha=1/3$ which corresponds 
to Simpson quadrature of the top-hat filter.
In actual simulations the resolved fields are known only on a set of 
grid points $\{x_{m}\}_{m=0}^{N}$. The application
of $L^{-1}$ to a general discrete solution $\{ \ub(x_{m})\}$ can be 
specified using discrete Fourier transformation as
\cite{kgvg99}
\be
L^{-1}(\ub_{m})={{\sum_{j=-n}^{n}}} \Big( 
\frac{\alpha-1+\sqrt{1-2\alpha}}{\alpha} \Big)^{|j|} 
\frac{{{\ub_{m+rj/2}}}}{(1-2\alpha)^{1/2}}
\en
where the subgrid resolution $r=\Delta/h$ is assumed to be even. An 
accurate and efficient inversion can be obtained with
only a few terms, recovering the original signal to within machine 
accuracy with $n \approx 10$. At fixed $\Delta$,
variation of the subgrid resolution $r$ allows an independent control 
over flow-smoothing and numerical representation
\cite{geurtsfroehlich2002}. Filtering and inversion in three dimensions arises from composing three one dimensional
filters.

To assess the Leray model the turbulent mixing layer is simulated in 
a volume $\ell^{3}$ at various $Re$ adopting a
fourth order accurate spatial discretization and explicit Runge-Kutta 
time-stepping. We compare predictions with those
obtained using the dynamic subgrid model, which was shown to be among 
the most accurate models in a comparative study of
the same turbulent mixing layer reported in~\cite{vreman_jfm}.

A first introductory test of the Leray model is obtained by studying 
instantaneous solutions. As a typical
illustration of the mixing layer the DNS prediction of the normal velocity
$u_{2}$ is shown in the
turbulent regime in Fig.~\ref{vortplot}(a). We used $Re=50$ based on 
the initial momentum thickness and free-stream
flow properties. The filtered $u_{2}$ can be seen in 
Fig.~\ref{vortplot}(b) establishing a significant smoothing
due to the `Simpson' filter at $\Delta=\ell/16$. The Leray prediction 
(Fig.~\ref{vortplot}(c)) appears to capture the main `character' as well as some of the details of
the filtered DNS solution. A slight
underprediction of the influence of the small scales is, however, 
apparent. Further visualization showed that the instantaneous Leray predictions display
much better overall agreement with filtered DNS than the dynamic model, which relative to the Leray model significantly
overpredicts the smoothing \cite{vreman_jfm}. Of course,
assessing the quality of LES predictions in this way is difficult to quantify and we
consider more specific measures next.

The evolution of a crucial mean-flow property such as the momentum 
thickness is shown in Fig.~\ref{fig1}. The Leray
results compare significantly better with filtered DNS results than 
those obtained with the dynamic model on $32^{3}$
grid-cells. We observe that some of the discrepancies between Leray 
and filtered DNS results are due to numerical
contamination. By increasing the resolution at fixed $\Delta$, a good 
impression of the grid-independent solution to
the modeled equations can be inferred using $64^3$ -- $96^3$ 
grid-cells, i.e., $\Delta/h=4$ to $6$
\cite{geurtsfroehlich2002}. Numerical contamination also plays a role 
in the dynamic model. The grid-independent
solution corresponding to the dynamic model appears less accurate 
than the corresponding Leray result.

A more detailed assessment is obtained from the streamwise kinetic 
energy spectrum shown in Fig.~\ref{fig2}. The dynamic
model yields a significant underprediction of the intermediate and 
smaller retained scales, particularly for the
approximately grid-independent solution. The Leray predictions are 
much better. On coarse grids, an overprediction of the
smaller scales is apparent due to interaction with the 
spatial discretization method. At proper numerical subgrid resolution the 
situation is considerably improved and the Leray model is
seen to capture all scales with high accuracy. A slight, systematic 
underprediction of the smaller scales remains,
consistent with the impression obtained from Figs.~\ref{vortplot}(b)-(c).

A particularly appealing property of Leray modeling is the robustness 
at very high Reynolds numbers, cf.
Fig.~\ref{fig3}. This is quite unique for a subgrid model without an 
explicit eddy-viscosity contribution. Although comparison
with filtered DNS data is impossible here, we observe that the 
smoothed Leray dynamics is essentially captured as
$r=\Delta/h \geq 4$ \cite{geurtsfroehlich2002}. The tail of the 
spectrum increases with $Re$, indicating a greater
importance of small scale flow features. Improved subgrid resolution 
shows a reduction of these smallest
scales, consistent with the reduced numerical error.
At high $Re$ the spectrum corresponding to the Leray model tends to contain a region with 
approximately $k^{-5/3}$ behavior, which is absent at $Re=50$.
Further analysis showed that the solution develops self-similarly at high $Re$.

The Leray model was presented to illustrate the new regularization 
approach for LES. It predicts mean flow
properties such as momentum thickness very accurately. The model 
exhibits both positive and negative production of
turbulent kinetic energy. The computational overhead associated with
the Leray model can be much lower than that of dynamic (mixed) 
models, especially if quantities are desired which are
rather insensitive to the inversion quality. The regularized Leray dynamics shows an appealing robustness at 
high $Re$. Further extensions of the regularization
approach are presently being considered. Of particular 
interest is the Lagrangian averaged NS$-\alpha$ model \cite{FHT[2001]} which arises in the 
Euler-Poincar\'e framework for smoothed flow
dynamics.

\section*{Acknowledgment}

This work developed from discussions the authors enjoyed while participating in the turbulence program
hosted by the Isaac Newton Institute in Cambridge (1999). BJG 
would like to acknowledge support from the Los
Alamos Turbulence Working Group and Center for Nonlinear Studies (2002).

\newpage

\begin{figure}

\caption{Normal velocity component $u_{2}$ at time $t=80$, (a): DNS, 
(b): filtered DNS, (c): Leray on $64^{3}$; using a
filterwidth $\Delta=\ell/16$. The light (dark) isosurfaces correspond 
to $u_{2}=0.3~(-0.3)$.}

\end{figure}

\begin{figure}
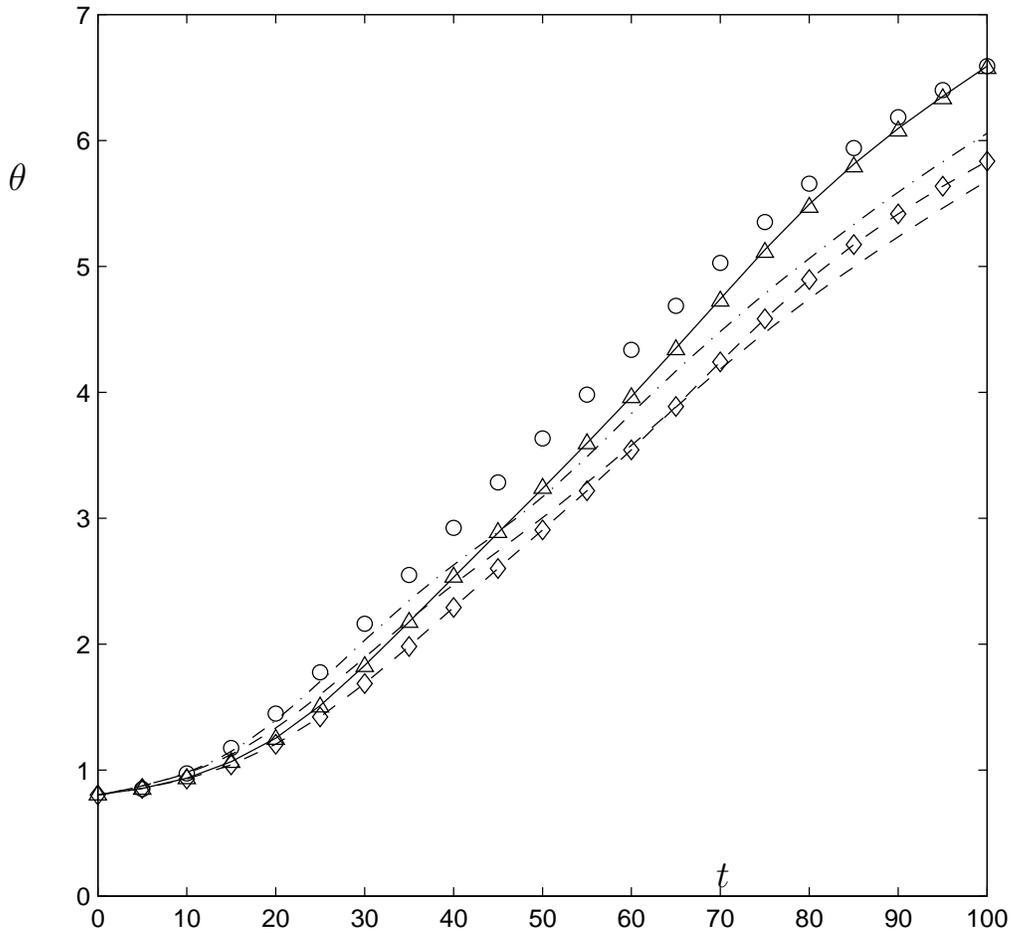


\caption{
Momentum thickness $\theta$: filtered DNS ($\circ$), Leray-model 
($32^3$: dash-dotted, $64^3$: solid, $96^3$:
$\triangle$), dynamic model ($32^3$: dashed, $64^3$: dashed with 
$\diamond$). A fixed filterwidth of $\ell/16$ was used.}

\end{figure}

\begin{figure}
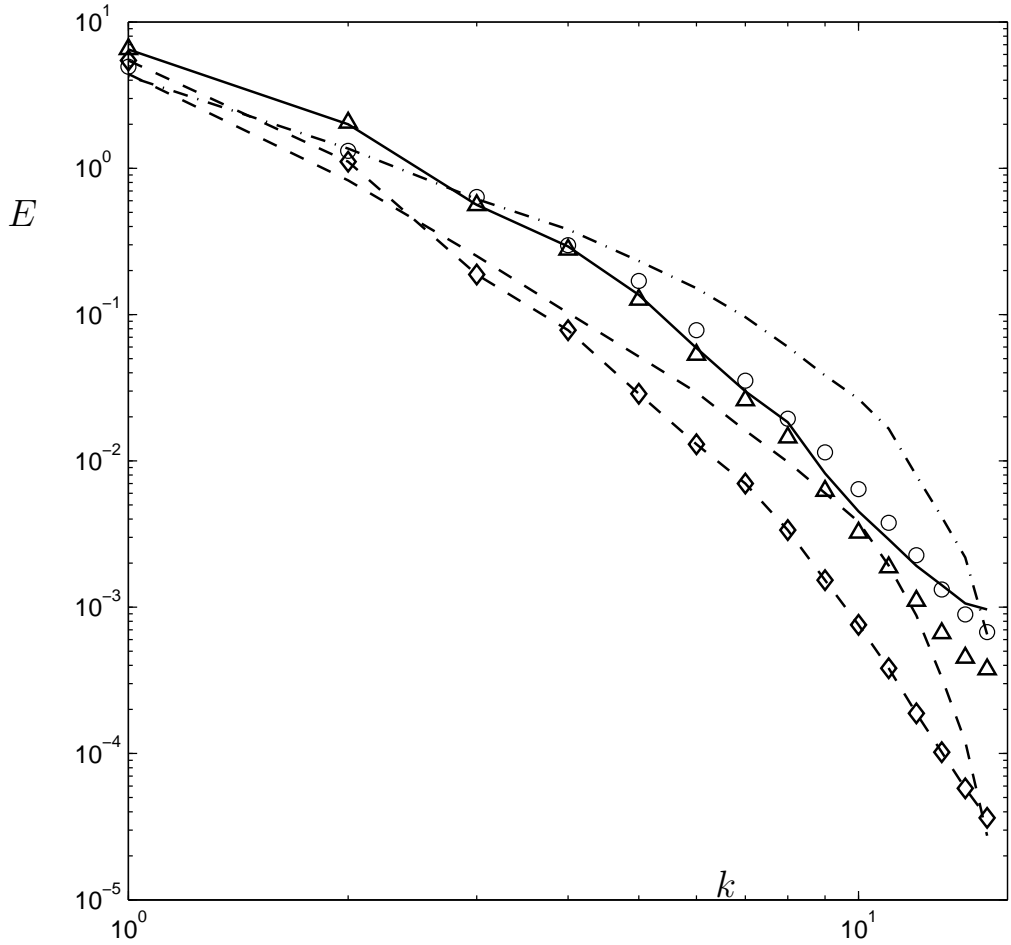


\caption{
Streamwise kinetic energy spectrum $E$ at $t=75$: filtered DNS 
($\circ$), Leray-model ($32^3$: dash-dotted,
$64^3$: solid, $96^3$: $\triangle$), dynamic model ($32^3$: dashed, 
$64^3$: dashed with $\diamond$). A fixed
filterwidth of $\ell/16$ was used.}

\end{figure}

\begin{figure}

\caption{
Streamwise kinetic energy spectrum $E$ at $t=75$ predicted by the 
Leray model: $Re=50$ ($64^3$: dash-dotted, $96^3$:
dash-dotted, $\triangle$), $Re=500$ ($64^3$: dashed, $96^3$: dashed, 
$\triangle$), $Re=5000$ ($64^{3}$: solid,
$96^3$: solid, $\triangle$). A fixed filterwidth of $\ell/16$ was 
used. The dotted line represents $k^{-5/3}$.}

\end{figure}
\newpage

\setcounter{figure}{0}

\begin{figure}[htb]

\centerline{
\psfig{figure=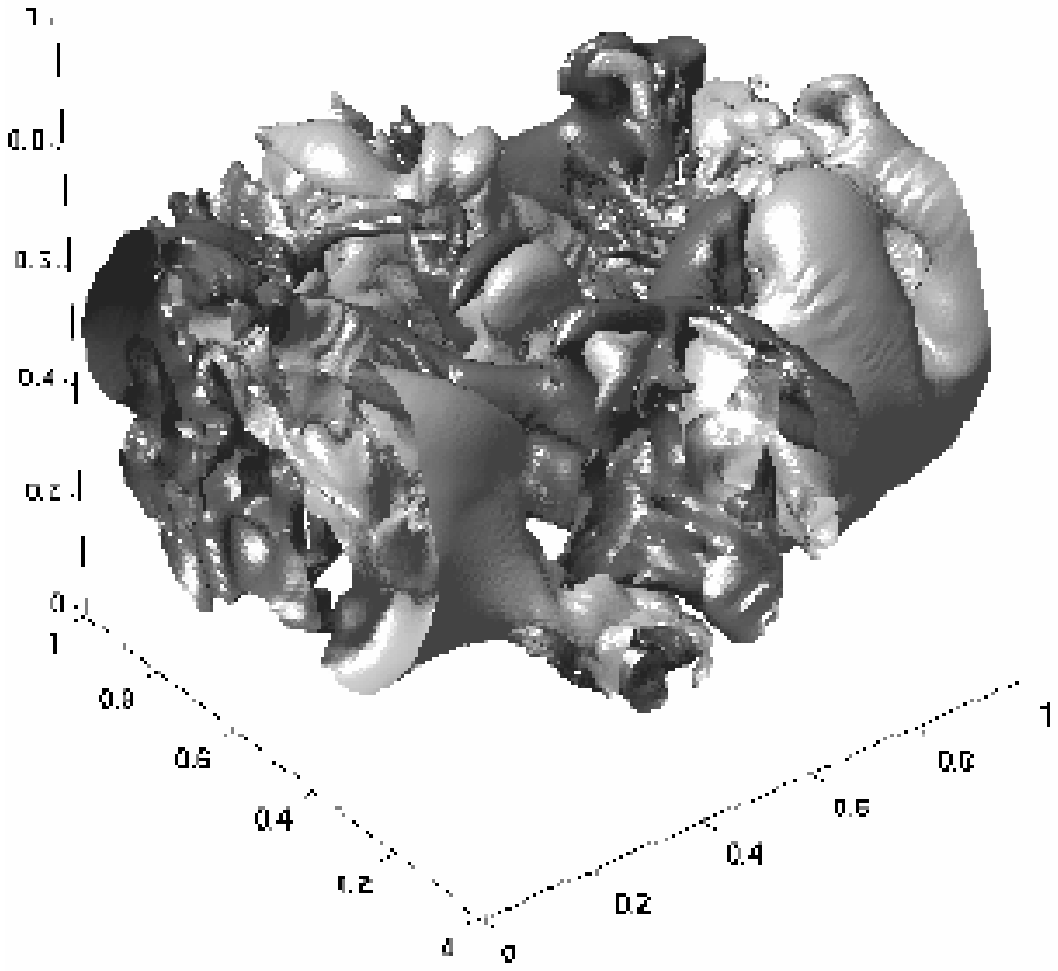,width=0.35\textwidth}}

\vspace*{-25mm}

\hfill{(a)}

\vspace*{-3mm}

\centerline{
\psfig{figure=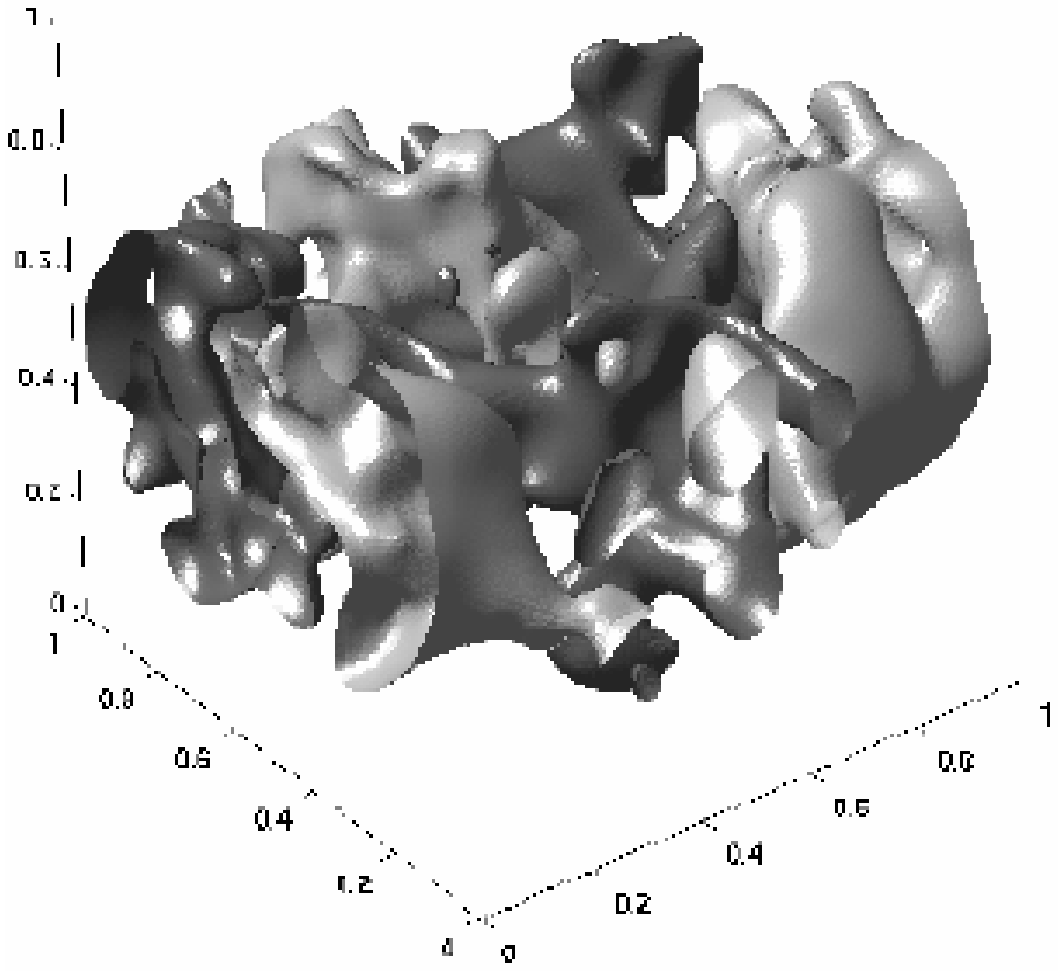,width=0.35\textwidth}}

\vspace*{-25mm}

\hfill{(b)}

\vspace*{-3mm}

\centerline{
\psfig{figure=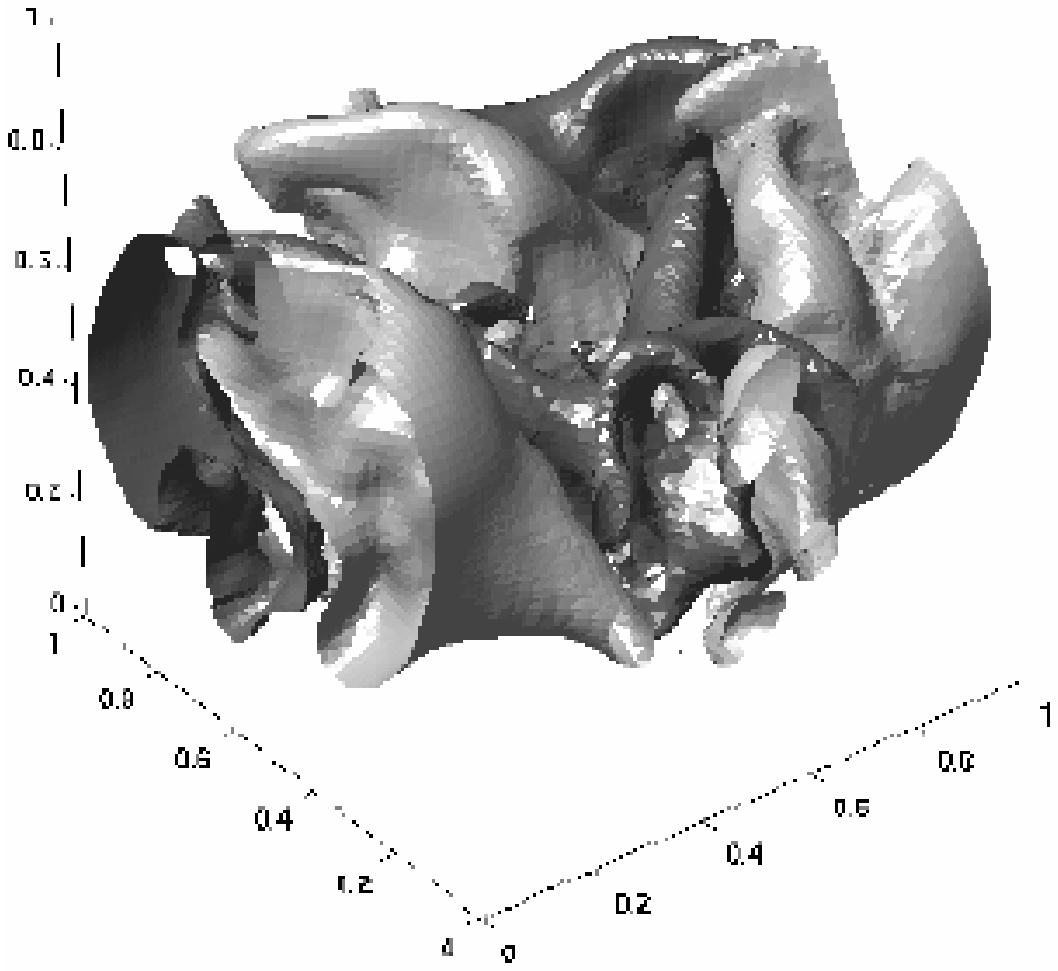,width=0.35\textwidth} }

\vspace*{-25mm}

\hfill{(c)}

\vspace*{15mm}

\caption{Normal velocity component $u_{2}$ at time $t=80$, (a): DNS, 
(b): filtered DNS, (c): Leray on $64^{3}$; using a
filterwidth $\Delta=\ell/16$. The light (dark) isosurfaces correspond 
to $u_{2}=0.3~(-0.3)$.}

\label{vortplot}
\end{figure}

\newpage

\begin{figure}[htb]

\centerline{
\psfig{figure=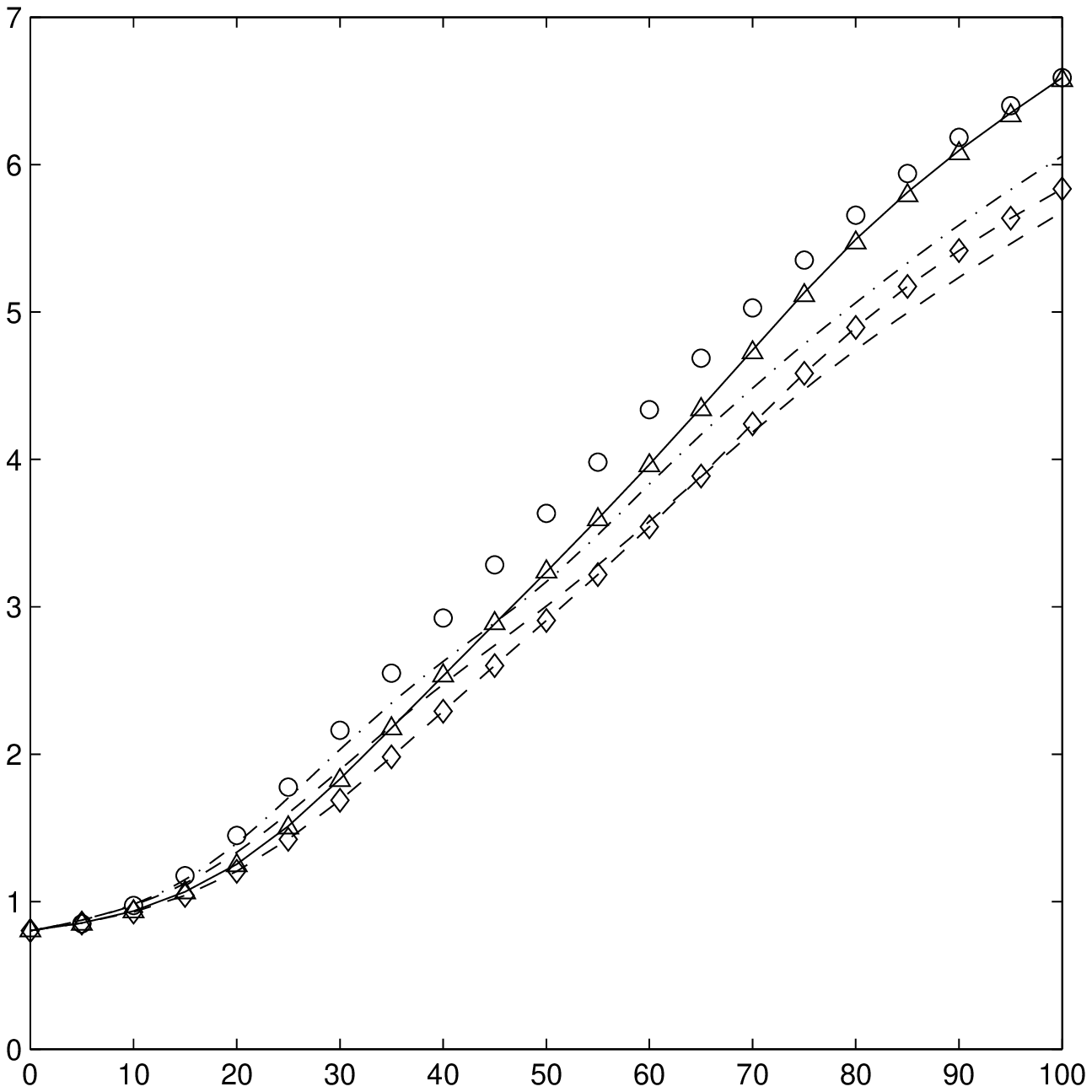,width=0.7\textwidth}
}

\vspace*{-0.575\textwidth}

\hspace*{0.075\textwidth} {\Large{$\theta$}}

\vspace*{0.495\textwidth}

\hspace*{0.6\textwidth} {\Large{$t$}}

\vspace*{10mm}

\caption{
Momentum thickness $\theta$: filtered DNS ($\circ$), Leray-model 
($32^3$: dash-dotted, $64^3$: solid, $96^3$:
$\triangle$), dynamic model ($32^3$: dashed, $64^3$: dashed with 
$\diamond$). A fixed filterwidth of $\ell/16$ was used.}

\label{fig1}
\end{figure}

\newpage

\begin{figure}[htb]

\centerline{
\psfig{figure=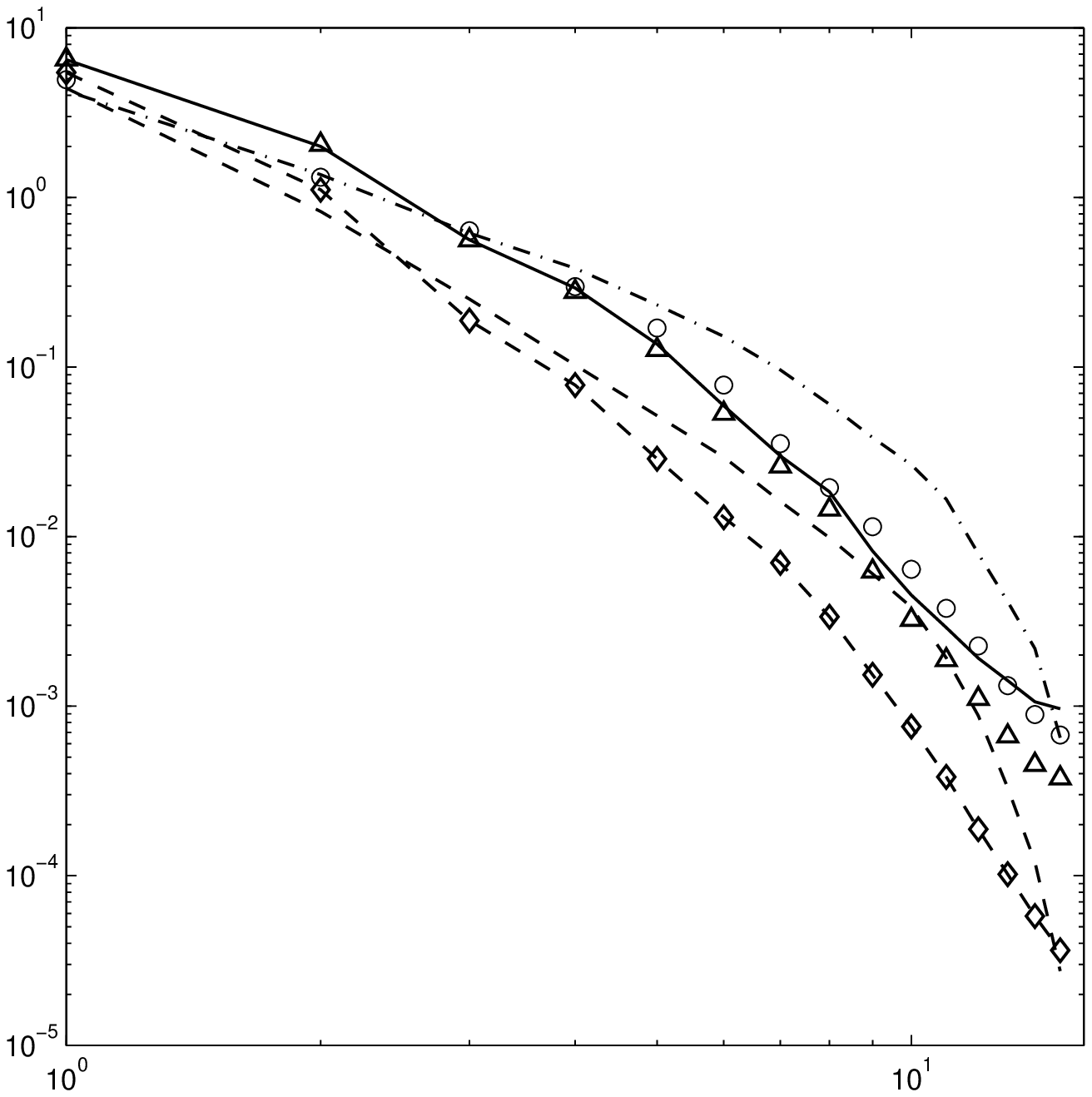,width=0.7\textwidth}
}

\vspace*{-0.555\textwidth}

\hspace*{0.075\textwidth} {\Large{$E$}}

\vspace*{0.475\textwidth}

\hspace*{0.6\textwidth} {\Large{$k$}}

\vspace*{10mm}

\caption{
Streamwise kinetic energy spectrum $E$ at $t=75$: filtered DNS 
($\circ$), Leray-model ($32^3$: dash-dotted,
$64^3$: solid, $96^3$: $\triangle$), dynamic model ($32^3$: dashed, 
$64^3$: dashed with $\diamond$). A fixed
filterwidth of $\ell/16$ was used.}

\label{fig2}
\end{figure}

\newpage

\begin{figure}[htb]

\centerline{
\psfig{figure=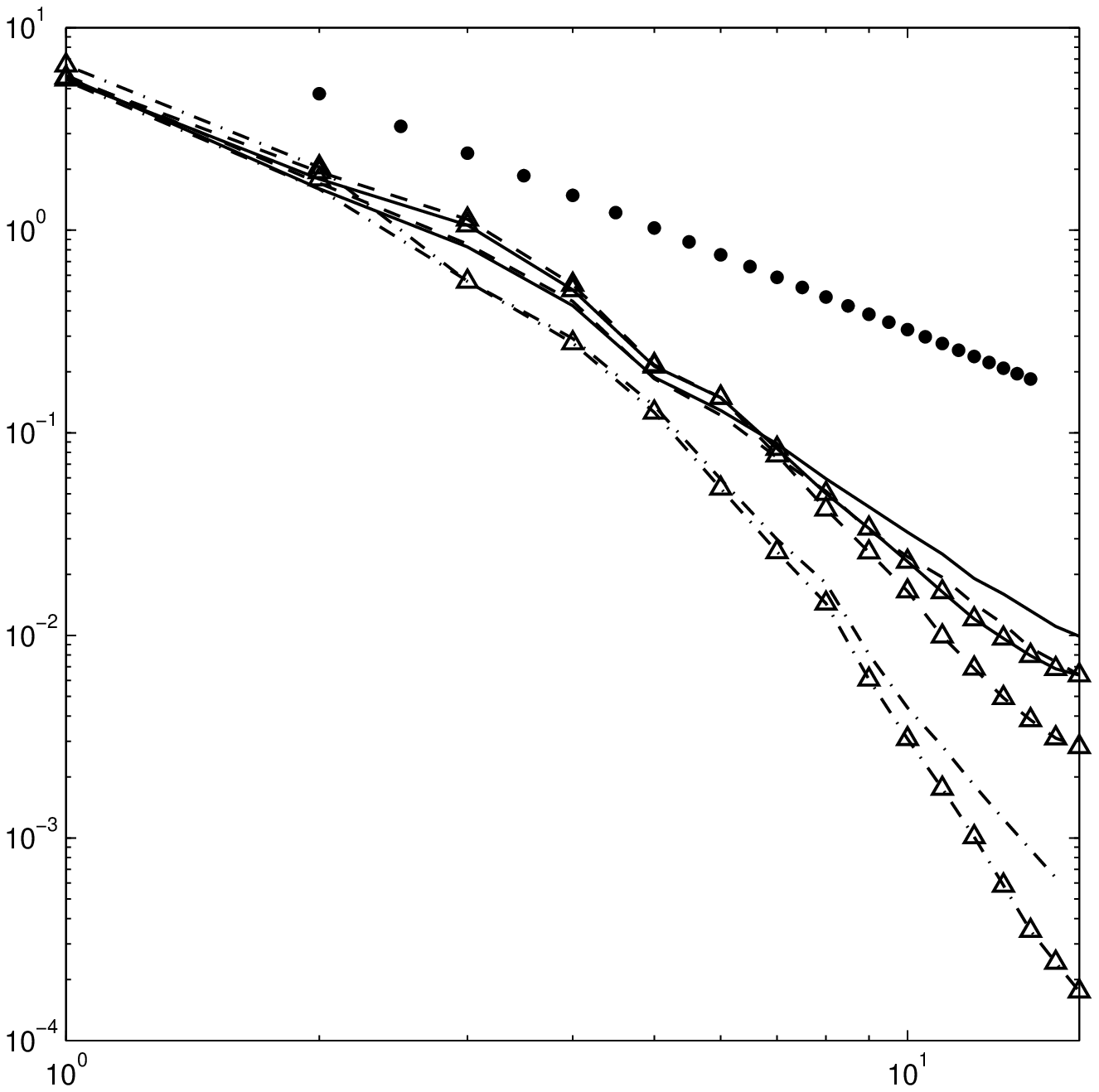,width=0.7\textwidth}
}

\vspace*{-0.525\textwidth}

\hspace*{0.075\textwidth} {\Large{$E$}}

\vspace*{0.45\textwidth}

\hspace*{0.6\textwidth} {\Large{$k$}}

\vspace*{10mm}

\caption{
Streamwise kinetic energy spectrum $E$ at $t=75$ predicted by the 
Leray model: $Re=50$ ($64^3$: dash-dotted, $96^3$:
dash-dotted, $\triangle$), $Re=500$ ($64^3$: dashed, $96^3$: dashed, 
$\triangle$), $Re=5000$ ($64^{3}$: solid,
$96^3$: solid, $\triangle$). A fixed filterwidth of $\ell/16$ was 
used. The dotted line represents $k^{-5/3}$.}

\label{fig3}
\end{figure}

\end{document}